\documentclass[aps,prl,reprint,groupedaddress]{revtex4-1}
\usepackage{graphicx} 
\usepackage{dcolumn} 
\usepackage{bm}
\usepackage{hyperref}

\begin{document}

\title{Correlation of the symmetry energy at subsaturation densities and neutron-skin thickness in low-energy antiproton induced reactions }
\author{Ban Zhang}
\author{Zhao-Qing Feng}
\email{Corresponding author: fengzhq@scut.edu.cn}
\affiliation{ School of Physics and Optoelectronics, South China University of Technology, Guangzhou 510640, China }

\begin{abstract}
Within the framework of Lanzhou quantum molecular dynamics transport model, the neutron-skin thickness and its impact on the nuclear dynamics induced by low-energy antiprotons are investigated thoroughly. The correlation of the neutron-skin thickness and stiffness of symmetry energy is implemented into the transport model via the Fermi distributions of the proton and neutron density profiles. It is found that antiprotons are predominantly annihilated in the subsaturation density region (0.4$\rho_{0}$-0.8$\rho_{0}$). The isospin ratios of free neutrons to protons (n/p) and charged pion yields ($\pi^{-}$/$\pi^{+}$) in collisions of antiprotons on $^{48}\rm{Ca}$ and $^{208}\rm{Pb}$ are analyzed systematically for extracting the symmetry energy in the domain of subsaturation densities. The n/p ratio is sensitive to the stiffness of symmetry energy in the low-density region and a soft symmetry energy leads to the larger n/p ratio, in particular with decreasing the beam momentum. The $\pi^-/\pi^+$ ratio is also enhanced with the soft symmetry energy at kinetic energies below 150 MeV.

\end{abstract}

\maketitle

\section{I. Introduction}

Understanding the properties of neutron-rich matter, particularly the behavior of symmetry energy at varying densities, is essential for advancing our knowledge of exotic nuclear structure, reaction dynamics and compact stars. The nuclear equation of state (EOS) is characterized by the binding energy per nucleon of nuclear matter, and the difference in the binding energy between pure neutron matter and symmetric nuclear matter is defined as the symmetry energy, which can be derived from the isospin dependent nucleon-nucleon force. The nuclear symmetry energy influences the structure of neutron-rich nuclei and stability of nuclear matter \cite{Ga24, Na19, Zhang13}, but is also crucial for the properties of compact stars, such as the supernova explosion, mass-radius relationship of neutron stars, pasta structure etc \cite{La01, La23, St05, Lin20}.

The neutron-skin is caused from the difference of neutron and proton density profiles of neutron-rich nuclei and manifests the symmetry energy properties in the density region of 0.1 fm$^{-3}$, which is defined as the difference between the root-mean-square (rms) radii of neutrons and protons, $\Delta r_{np} = \left\langle r_n^2 \right\rangle^{\frac{1}{2}} - \left\langle r_p^2 \right\rangle^{\frac{1}{2}}$. The neutron-skin thickness is related to the collective multiple interactions inside a nucleus and symmetry energy. The neutron-skin thickness of $^{48}$Ca and $^{208}$Pb has been investigated by the nuclear structure models, such as the relativistic mean-field model, Skyrme Hartree-Fock method etc \cite{Br00,Ty01,Yo04,Ce09,Re10,Ro11,Ag12,Ho01,Fu02,Yo06,To05,Wa09,Ro14,Ag12,Oy03,Da03,Ai87}. The magnitude of thickness might be extracted in experiments from the giant dipole resonance (GDR) \cite{Kr91, Cs03}, spin dipole resonance (SDR) \cite{Kr99,Ga81}, The X-ray cascade from antiprotonic atoms \cite{Lu94,Lu98,Sc98,Tr01,Kl07}, proton elastic-scattering (PES) \cite{Te08,Ze10,Ta11}, parity violating in electron-nucleus scattering (PVES) \cite{Ab12,Ad21,Ad22}, reaction cross section measurements in heavy-ion collisions \cite{Ta22} etc.

In 1955, the first evidence of antiprotons was discovered at Berkeley national laboratory (BNL) by Chamberlain \textit{et al}.\cite{Ch55}. As a secondary beam, the antiproton induced nuclear reactions have been extensively investigated in different laboratories in the world, such as the low-energy antiproton ring (LEAR) at CERN \cite{Go96}, the national laboratory for high-energy physics at KEK \cite{Mi84}, and the BNL alternating gradient synchrotron accelerator \cite{Le99}, etc. Several models have been established for describing the experimental phenomena, such as the intranuclear cascade model \cite{Cu89}, the statistical approach \cite{Go92,Bo95}, the Giessen Boltzmann-Uehling-Uhlenbeck transport model \cite{La12} and the Lanzhou quantum molecular dynamics (LQMD) approach \cite{feng14}. The elastic scattering and annihilation reactions have the similar magnitude of cross sections. The main products in the annihilation of an antiproton and a proton (neutron) are the energetic pions. The annihilation cross section of the low-energy antiproton with a proton is approximately twice that of the annihilation with a neutron \cite{Ba87}. A method to study the nuclear surface from the low-energy antiproton annihilation was proposed by Wycech and Piscichia \cite{Wy23}. It has been reported that the neutron-skin thickness of $\Delta r_{np} = 0.16 \pm (0.02)_{stat} \pm (0.04)_{syst}$ fm for $^{208}$Pb was obtained by the X-ray cascade from antiprotonic atoms \cite{Kl07}. The measurements of the neutron-skin thickness via the elastic scattering and annihilation reactions between antiprotons and nucleons are to be expected in the future experiments.

In this work, the correlation of neutron-skin thickness and symmetry energy in the low-energy antiproton induced reactions on $^{48}\rm{Ca}$ and $^{208}\rm{Pb}$ is thoroughly investigated for the first time within the framework of LQMD transport model. The article is organized as follows. Section \uppercase\expandafter{\romannumeral2} is a brief introduction of the model. The experimental observables for extracting the neutron-skin thickness in the antiproton induced reactions are discussed in section \uppercase\expandafter{\romannumeral3}. The perspective on the measurements of the neutron-skin thickness via the low-energy antiproton induced reactions is summarized in \uppercase\expandafter{\romannumeral4}.

\section{II. Model description}

The LQMD transport model has been used for heavy-ion collisions and hadron (proton, antiproton, meson, hyperon etc) induced reactions, in which the elastic and inelastic hadron-hadron collisions, the production, decay and reabsorption of resonances, the interaction potential between hadrons and nucleons are self-consistently implemented into the model \cite{feng11}. The model has been used for describing the nuclear dynamics induced by antiproton, proton, meson and hyperon, nuclear fragmentation and hypernuclide production, dense matter properties via heavy-ion collisions etc \cite{feng12, liu23}. The particle production, nuclear fragmentation reaction and hypernuclear production in the antiproton induced reactions have been thoroughly investigated, i.e., the phase-space distributions and yields of pions, kaons, hyperons, nuclear and hypernuclear fragments \cite{feng14, feng16}. The correlation of the symmetry energy at subsaturation densities and neutron-skin thickness is to be investigated in the low-energy antiproton induced reactions for the first time.

\subsection{1. Mean-field potentials}
In the LQMD model, each nucleon of reaction system is represented by a Gaussian wave-packet as
\begin{equation}
\phi_i(\textbf{r},t) = \frac{1}{(2\pi L)^\frac{3}{4}} \exp\left[ -\frac{(\textbf{r}-\textbf{r}_i(t))^2}{4L} + \frac{i\textbf{p}_i(t)\cdot\textbf{r}}{\hbar}\right],
\end{equation}
where the $L$ represents the square of the wave-packet width and is relative to the mass number of projectile or target nucleus as $\sqrt{L}=0.08 A^{\frac{1}{3}}+0.99$ fm. The variables $\textbf{r}_i(t)$ and $\textbf{p}_i(t) $ denote the center position of the $i-$th wave-packet in the coordinate and momentum space, respectively. The total wave-function of the reaction system is the direct product of each nucleon wave-function by
\begin{equation}
\Phi_i(\textbf{r},t)=\prod_{i}\phi_i(\textbf{r},\textbf{r}_i,\textbf{p}_i, t).
\end{equation}

The temporal evolution of nucleons and resonances is governed by the self-consistently generated mean-field potentials as
\begin{equation}
\dot{\textbf{r}_{i} }  =  \frac{\partial H}{\partial \textbf{p} _{i} }, \quad
\dot{\textbf{p}_{i} } =  -\frac{\partial H}{\partial \textbf{r} _{i} }.
\end{equation}
The Hamiltonian consists of the relativistic energy, Coulomb interaction, and local interaction as follows
\begin{equation}
H_{B}=\sum_{i}\sqrt{\textbf{p}_{i}^{2}+m_{i}^{2}}+U_{Coul}+U_{loc}.
\end{equation}
Here the $\textbf{p}_{i}$ and $m_{i}$ represent the momentum and the mass of the baryons. The local interaction potential is evaluated by the energy-density functional as
\begin{equation}
U_{loc}=\int V_{loc}[\rho(\bm{r})]d\bm{r}
\end{equation}
with
\begin{eqnarray}
 V_{loc} && = \frac{\alpha}{2}\frac{\rho^2}{\rho_0} + \frac{\beta}{1+\gamma}\frac{\rho^{1+\gamma}}{\rho_0^\gamma}+E_{sym}^{loc}(\rho)\rho\delta^2    
 \nonumber\\
&& + \frac{g_{sur}}{2\rho_0}(\nabla \rho)^2+\frac{g_{sur}^{iso}}{2\rho_0}[\nabla (\rho_n-\rho_p)]^2
\end{eqnarray}
with the isospin asymmetry $\delta=(\rho_n-\rho_p)/(\rho_n + \rho_p)$, the neutron density $ \rho_n$ and proton density $\rho_p$, respectively. The nuclear matter parameters with $\alpha = -226.5$ MeV, $\beta = 173.7$ MeV and $\gamma = 1.039$ lead to the incompressibility modulus of $ K = 230$ MeV of the isospin symmetry nuclear matter at the saturation density ($\rho_0 = 0.16\ fm^{-3}$). The surface coefficients $g_{sur}$ and $g_{sur}^{iso}$ are taken as 23 MeV fm$^{2}$ and -2.7 MeV fm$^{2}$ for describing the binding energies of finite nuclei.

The symmetry energy is composed of the kinetic energy from the nucleonic Fermi motion and the local density dependent term as
\begin{equation}
E_{sym}(\rho) = \frac{1}{3}\frac{\hbar^{2}}{2m}(\frac{3}{2}\pi^{2}\rho)^{2/3} + \frac{1}{2} C_{sym}(\rho /\rho _{0} )^{\gamma _{s}}.
\end{equation}
The coefficient $C_{sym}$ is taken to be 38 MeV, which leads to the symmetry energy of 31.5 MeV at the saturation density. The stiffness parameter $\gamma_s$ is adjusted for getting the density dependence of symmetry energy, e.g., the values of 0.5, 1, and 2 being the soft, linear, and hard symmetry energy, corresponding to the slope parameters $[L(\rho _{0}) = 3\rho_{0} dE_{sym}(\rho)/d\rho|_{\rho=\rho_{0}}]$ of 53, 82, and 139 MeV, respectively. The slope parameter is closely related to the mass-radius relation of neutron stars and might be extracted from the antiproton-nucleus collisions.

The one-body potential for the antiproton transportation in the nuclear medium is calculated by performing the G-parity transformation of nucleon self-energies. The optical potential of antiproton in nuclear medium is derived from the in-medium energy as
\begin{equation}
V_{\overline{p}}(\textbf{p},\rho)=\omega_{\overline{N}}(\textbf{p},\rho)-\sqrt{\textbf{p}^{2}+m^{2}}.
\end{equation}
The antinucleon energy in nuclear medium is evaluated by the dispersion relation as
\begin{equation}
\omega_{\overline{N}}(\textbf{p}_{i},\rho_{i})=\sqrt{(m_{N}+\Sigma_{S}^{\overline{N}})^{2}+\textbf{p}_{i}^{2}} + \Sigma_{V}^{\overline{N}}
\end{equation}
with $\Sigma_{S}^{\overline{N}}=\xi\Sigma_{S}^{N}$ and $\Sigma_{V}^{\overline{N}}=-\xi\Sigma_{V}^{N}$ with $\xi$=0.25. The strength of the optical potential $V_{\overline{N}}=-164$ MeV is obtained at the normal nuclear density $\rho_{0}$=0.16 fm$^{-3}$ by fitting the available experimental data of antiproton-nucleus scattering \cite{La12, feng14}.

\subsection{2. Correlation of initialization and symmetry energy}
The neutron-skin thickness $r_{skin}=r_{n}-r_{p}$ is correlated with the slope parameter $L$ of symmetry energy. Since the symmetry energy cannot be measured directly, its slope is inferred experimentally from the determination of the neutron skin thickness. The high-resolution $E$1 polarizability experiment$(E1pE)$ conducted at the Research Center for Nuclear Physics (RCNP) on $^{48}\rm{Ca}$ \cite{Bi16}and $^{208}\rm{Pb}$ \cite{Ta11} have yielded neutron skin thickness estimates of $ r_{skin}^{48}(E1pE) = 0.14-0.20 $ fm, $ r_{skin}^{208}(E1pE)=0.135-0.181$ fm. Analyses from the proton \cite{Ze10} and pion \cite{Fr12} nucleus scattering provided the neutron-skin thicknesses of $0.211\pm 0.06$ fm and $0.16\pm 0.07$ fm for $^{208}\rm{Pb}$, respectively. Studies of the annihilation of antiprotons on the nuclear surface \cite{Kl07, Br07} reported $r_{skin}^{208}=0.18 \pm0.04 ({expt.}) \pm0.05 ({theor.}) $ fm, considering both experimental and theoretical uncertainties. Isospin diffusion in heavy-ion collisions have determined $r_{skin}^{208}=0.22\pm0.04$ fm \cite{Chen05}. Tanaka and colleagues have determined the neutron skin thickness across calcium isotopes using interaction cross-section data\cite{Tanaka19}. Specifically for $^{48}\rm{Ca}$, the findings indicate a neutron skin thickness of $ r_{skin}^{48} = 0.146\pm0.048 $ fm. Tagami et al. have recently utilized the Gogny-D1S Hartree-Fock-Bogoliubov model to estimate the neutron skin thickness, reporting a value within the range of $ r_{skin}^{48}= 0.159-0.190$ fm \cite{Tagami19}. Following the first experimental outcome of the Lead Radius Experiment PREX-I experiment \cite{Ab12}, which reported a neutron skin thickness of $r_{skin}^{208}=0.33_{-0.18}^{+0.16}$ fm. Recently, the Lead Radius Experiment (PREX) collaboration has obtained the PREX-II results\cite{Ad21}. The reported neutron skin thickness value is $0.283\pm0.071$ fm. Furthermore, the Calcium Radius Experiment (CREX) \cite{Ad22} for $^{48}\rm{Ca}$ has yielded a precise measurement of the neutron-skin thickness, with a value for $ r_{skin}^{48}$ ranging from $0.071 - 0.171$ fm. In this work, for $^{208}\rm{Pb}$, we have selected neutron skin thickness values of 0.18 fm and 0.38 fm within the range of results from antiproton annihilation measurements and  the thicker neutron skin, respectively. Furthermore, to examine $^{48}\rm{Ca}$ , we have chosen values of 0.08 fm and 0.16 fm from within the measurement range of the CREX. These values correspond to the soft and hard symmetry energies within the LQMD model,  the symmetry energy slopes for  $^{48}\rm{Ca}$ and $^{208}\rm{Pb}$ are found to exhibit a linear relationship with the neutron-skin thicknesses, specifically characterized by the equations
$L=1075{r_{skin}^{48}-33}$ MeV for Ca and $L = 430{r_{skin}^{208}} - 24.4$ MeV for Pb, which exists a discrepancy from the linear fitting of nuclear structure models with $L=680{r_{skin}^{208}-68.7}$ MeV \cite{Ro11}. The increasing trend of the neutron-skin thicknesses with the slope parameter is similar. In this work, the proton and neutron density distributions of target nuclei is calculated by the well-known Skyrme Hartree-Fock model with the SKM* parameter, following which the distribution is reappeared using a two-parameter Fermi distribution
\begin{equation}
\rho^{T}_i=\frac{\rho^{T}_{0i}}{1+\exp(\frac{r-R^{T}_i}{a_i})}, \ i=n,p\ ,
\end{equation}
where $\rho^{T}_{0n(p)}$ is the central density of neutrons (protons) in the target nucleus, $R^{T}_{n(p)}$, $a_{n(p)}$ are the radius and diffusion coefficient of neutrons (protons) density distributions, respectively. In the system, it is taken into account that the number of particles is conserved, $N=\int \rho_n(\bm{r})d^3r$ and $Z=\int \rho_p(\bm{r})d^3r$. We fixed the proton density distribution while varying the diffusion coefficient of the neutron density distribution, an increased diffusion coefficient results in a larger rms radius for neutron, thereby enhancing the neutron skin thickness. For $^{208}\rm{Pb}$ and $ r_{skin}^{48}$, both distributions are parametrized with the parameters given in Table \ref{tab:1}.

\begin{table}[t]
\centering
\caption{Parameters of the Fermi distribution for $^{208}\rm{Pb}$ and $^{48}\rm{Ca}$, respectively.}
\label{tab:1}
\renewcommand{\arraystretch}{1.5}
\begin{tabular}{cccccccc}
\hline
\hline
 & \(p_{0p} (\text{fm}^{-3})\) &\(p_{0n} (\text{fm}^{-3})\)& \(R_p (\text{fm})\) & \(R_n (\text{fm})\) & $a_p (\text{fm})$ & $a_n (\text{fm})$ & $L (\text{MeV})$ \\[0.5ex]
\hline
$^{48}\rm{Ca}$ & 0.0731 & 0.0872 & 3.790 & 4.062 & 0.54 & 0.49 & 53\\[0.5ex]
                             & 0.0731 & 0.0854 & 3.790 & 4.062 & 0.54 & 0.53 & 139\\[0.5ex]
$^{208}\rm{Pb}$ & 0.0633 & 0.090 & 6.638 & 6.780 & 0.506 & 0.57 & 53\\[0.5ex]
                               & 0.0633 & 0.090 & 6.638 & 6.730 & 0.506 & 0.66 & 139\\[0.5ex]
\hline
\hline
\end{tabular}
\end{table}

The coordinates of nucleons are derived from uniform sampling of the nuclear density distribution, ensuring the discrepancy between the sampled rms (root-mean-square) radius and the initial rms radius is less than 0.1 fm, which results in a uniform and stable nucleon distribution. The Fermi momentum is calculated using the formula $P_F^i(\bm{r})= \hbar [3\pi^2\rho_i(\bm{r})]^\frac{1}{3}$, $i=n,p$, and the momentum coordinates are derived from random sampling within the range [0, $P_F^i$]. Subsequent sampling is performed until the difference between the nuclear binding energy and its experimentally determined value falls within an acceptable range. The stability of the initial nucleus is assessed and confirmed to be sustainable up to 300 fm/c under mean-field evolution. Fluctuations are observed in the rms radii and neutron skin thicknesses derived from this methodology. However, these variations remain within the desired limits. The figure 1 presents the rms radii of neutrons and protons, as well as the neutron skin, for $^{48}\rm{Ca}$ and $^{208}\rm{Pb}$, respectively. It is evident that the rms radius of protons remains essentially constant across various neutron skin thickness settings. Even under minor fluctuations, a distinct neutron skin thickness is sustained up to a time evolution of 300 fm/c.

\begin{figure*}
\centering
\includegraphics[width=\linewidth]{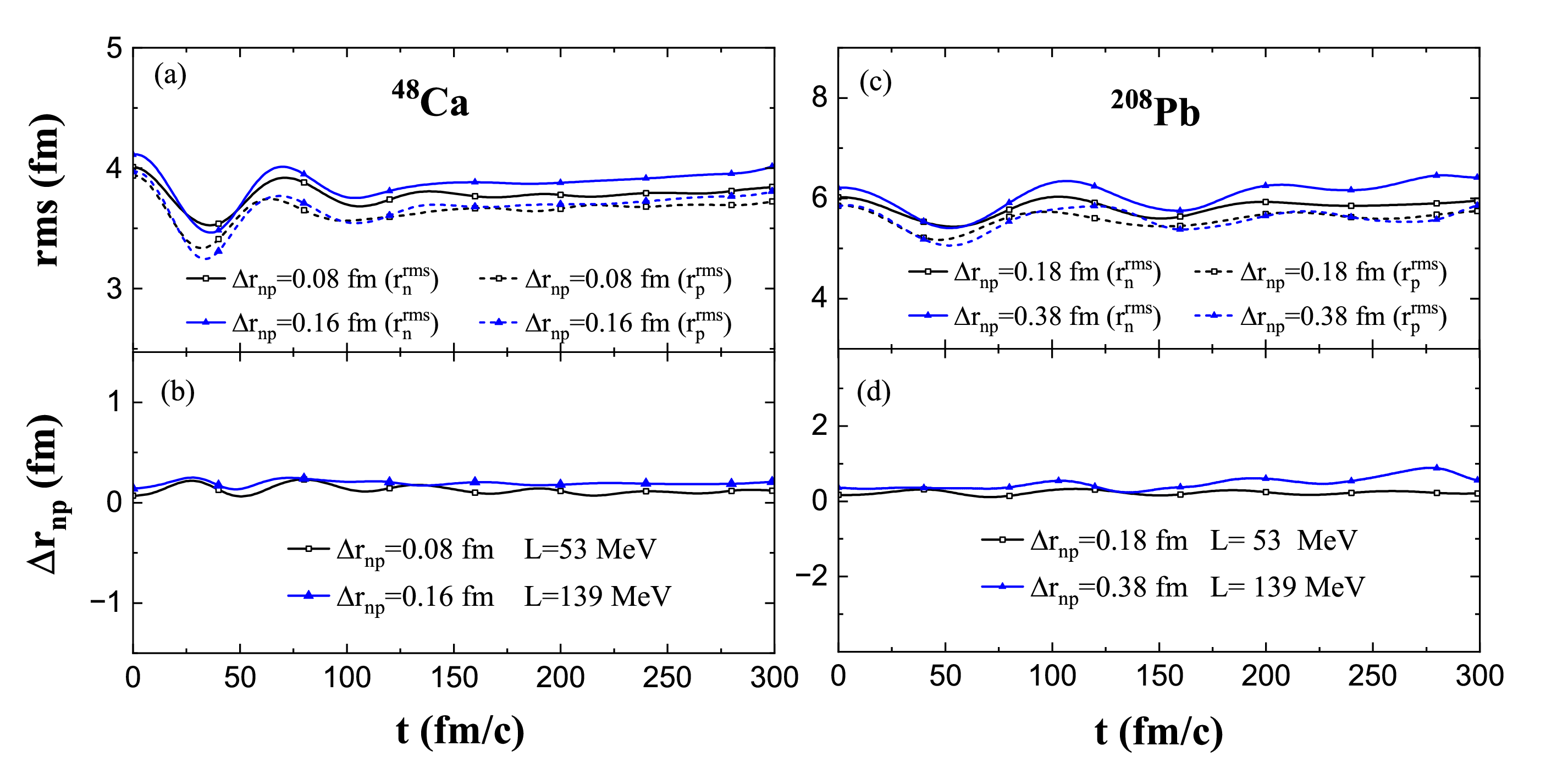}
\caption{\label{1} The time evolution of rms radii and neutron-skin thicknesses with different initial values for $^{48}\rm{Ca}$ (left panel) and $^{208}\rm{Pb}$ (right panel), respectively. }
\end{figure*}

\subsection{3. Reaction channels induced by antiprotons}
To investigate the antiproton-nucleus dynamics, the annihilation channels, charge-exchange reactions (CEX), elastic (EL) and inelastic scattering between the antiproton and nucleon collisions are included as follows \cite{feng16},
\begin{eqnarray}
&&  \overline{B}B \rightarrow \texttt{annihilation}(\pi,\eta,\rho,\omega,K,\overline{K},\eta\prime,K^{\ast},\overline{K}^{\ast},\phi),
\nonumber  \\
&&  \overline{B}B \leftrightarrow \overline{B}B (\texttt{CEX, EL}),   \overline{N}N \leftrightarrow \overline{N}\Delta(\overline{\Delta}N), \overline{B}B \leftrightarrow \overline{Y}Y,        \nonumber  \\
&&  \overline{B}B \leftrightarrow \overline{\Xi}\Xi, \overline{K}B \leftrightarrow K\Xi, YY \leftrightarrow N\Xi, \overline{K}Y \leftrightarrow \pi\Xi
\end{eqnarray}
Here the $B$ strands for a nucleon and $\Delta$(1232), Y($\Lambda$, $\Sigma$), $\Xi(\Xi^{0,-})$, K(K$^{0}$, K$^{+}$) and $\overline{K}$($\overline{K^{0}}$, K$^{-}$). The overline of B (Y) means its antiparticle. Mesons are the main products in the antiproton induced reactions. Besides the strangeness exchange reactions such as $\overline{K}N \rightarrow \pi Y$, hyperons are also contributed from the meson induced reactions $B\pi(\eta) \leftrightarrow YK$.

The main products in the antiproton annihilation reactions are the pions, which manifest a broad energy distribution and interact with the surrounding nucleons. The cross section of pion-nucleon scattering is evaluated with the Breit-Wigner formula as the form of
\begin{eqnarray}
\sigma_{\pi N\rightarrow R}(\sqrt{s}) && = \sigma_{\mathrm{max}}(\textbf{p}_{0}/\textbf{p})^{2}     \nonumber \\
&& \times\frac{0.25\Gamma^{2}(\textbf{p})}{0.25\Gamma^{2}(\textbf{p})+(\sqrt{s}-m_{0})^{2}},
\end{eqnarray}
where the $\textbf{p}$ and $\textbf{p}_{0}$ are the momenta of pions at the energies of $\sqrt{s}$ and $m_{0}$, respectively, and $m_{0}$ being the centroid of resonance mass, e.g., 1.232 GeV, 1.44 GeV and 1.535 GeV for $\Delta$(1232), $N^{\ast}$(1440), and $N^{\ast}$(1535), respectively. The maximum cross section $\sigma_{max}$ is taken from fitting the total cross sections of the available experimental data in pion-nucleon scattering with the Breit-Wigner form of resonance formation \cite{Li01}. For example, 200 mb, 133.3 mb, and 66.7 mb for $\pi^{+}+p\rightarrow \Delta^{++}$ ($\pi^{-}+n\rightarrow \Delta^{-}$), $\pi^{0}+p\rightarrow \Delta^{+}$ ($\pi^{0}+n\rightarrow \Delta^{0}$) and $\pi^{-}+p\rightarrow \Delta^{0}$ ($\pi^{+}+n\rightarrow \Delta^{+}$), respectively. And 24 mb, 12 mb, 32 mb, 16 mb for $\pi^{-}+p\rightarrow N^{\ast 0}(1440)$ ($\pi^{+}+n\rightarrow N^{\ast +}(1440)$), $\pi^{0}+p\rightarrow N^{\ast +}(1440)$ ($\pi^{0}+n\rightarrow N^{\ast 0}(1440)$), $\pi^{-}+p\rightarrow N^{\ast 0}(1535)$ ($\pi^{+}+n\rightarrow N^{\ast +}(1535)$) and $\pi^{0}+p\rightarrow N^{\ast +}(1535)$ ($\pi^{0}+n\rightarrow N^{\ast 0}(1535)$), respectively.

\section{III. Results and discussion}
In this work, the LQMD model is used to investigate the antiproton induced reactions on the neutron-rich nuclei $^{48}\rm{Ca}$ and $^{208}\rm{Pb}$. The antiproton-induced reactions provide significant insights for the energy deposition in the nuclear medium, the in-medium properties of antiproton-nucleon, meson-nucleon and hyperon-nucleon scatterings. The isospin ratios of neutron/proton (n/p) and $\pi^-$/$\pi^+$ ratio produced in the low-energy antiproton induced reactions are to be investigated for extracting the neutron-skin thickness and symmetry energy in the low-density region. The initial distance for the reactions is chosen from the antiproton and the surface distance of target nucleus with the value of 20 fm. We integrated the collision centrality (impact parameter) to the target radius in the antiproton induced reactions.

\begin{figure}[ht]
\centering
\includegraphics[width=\linewidth]{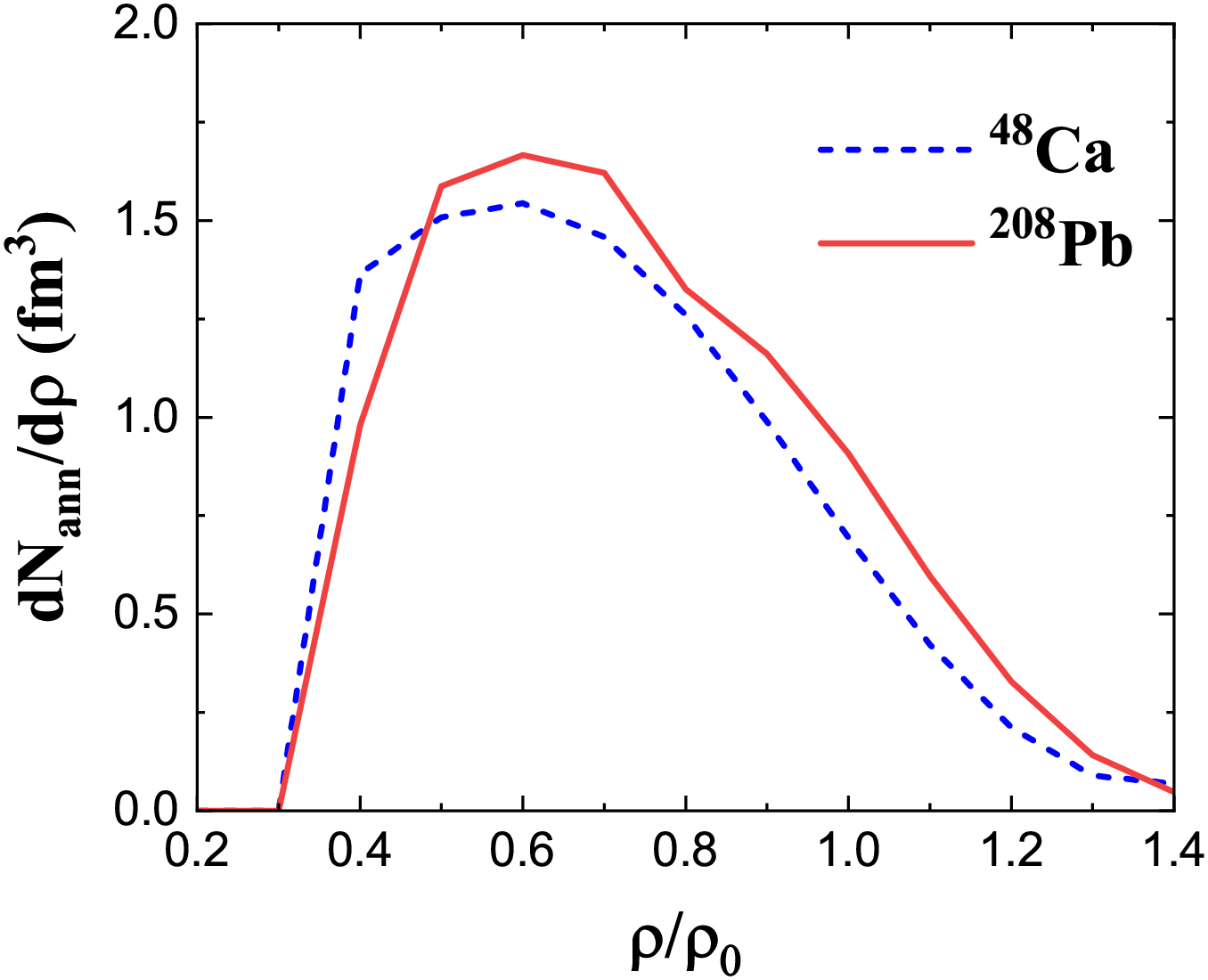}
\caption{\label{2} The density profiles at the annihilation of an antiprotons on $^{48}\rm{Ca}$ and $^{208}\rm{Pb}$ at the incident momentum of 200 MeV/c.}
\end{figure}

Within the framework of LQMD transport model, the nuclear fragmentation in the antiproton induced reactions is to be investigated systematically. The fragments are recognized by the well-known coalescence model with the minimum spanning tree (MST) method, in which the nucleons within the relative distance $\rm{R}_0$ (3 fm) and relative momentum $\rm{P}_0$ (200 MeV/c) are included into a fragment with the checking the root-mean-square radii \cite{feng17}. Compared to the proton-induced reactions or heavy-ion collisions \cite{Co16,feng16_2}, the antiproton-nucleon annihilation releases an immense concentration of energy \cite{Ra80}, which enables the energetic nucleons and mesons emission. The annihilation process of an antiproton primarily occurs on the nuclear surface \cite{Le74}, and the cross section for the low-velocity antiproton-proton annihilation is approximately twice in comparison with the antiproton-neutron annihilation \cite{Ba87}. In the neutron-rich nuclei, the appearance of the neutron-skin enhances the annihilation probability with neutrons \cite{Bu73}. The density profiles of an antiproton annihilation on $^{48}\rm{Ca}$ and $^{208}\rm{Pb}$ are shown in Fig. 2. It is clearly noticed that the antiproton predominantly annihilates in the subsaturation density region (0.4$\rho_0$-0.8$\rho_0$), which enables the available probes for extracting the low-density symmetry energy and neutron-skin thickness, e.g. free neutron/proton and $\pi^-/\pi^+$ in the antiproton induced reactions.

\begin{figure}[ht]
\centering
\includegraphics[width=\linewidth]{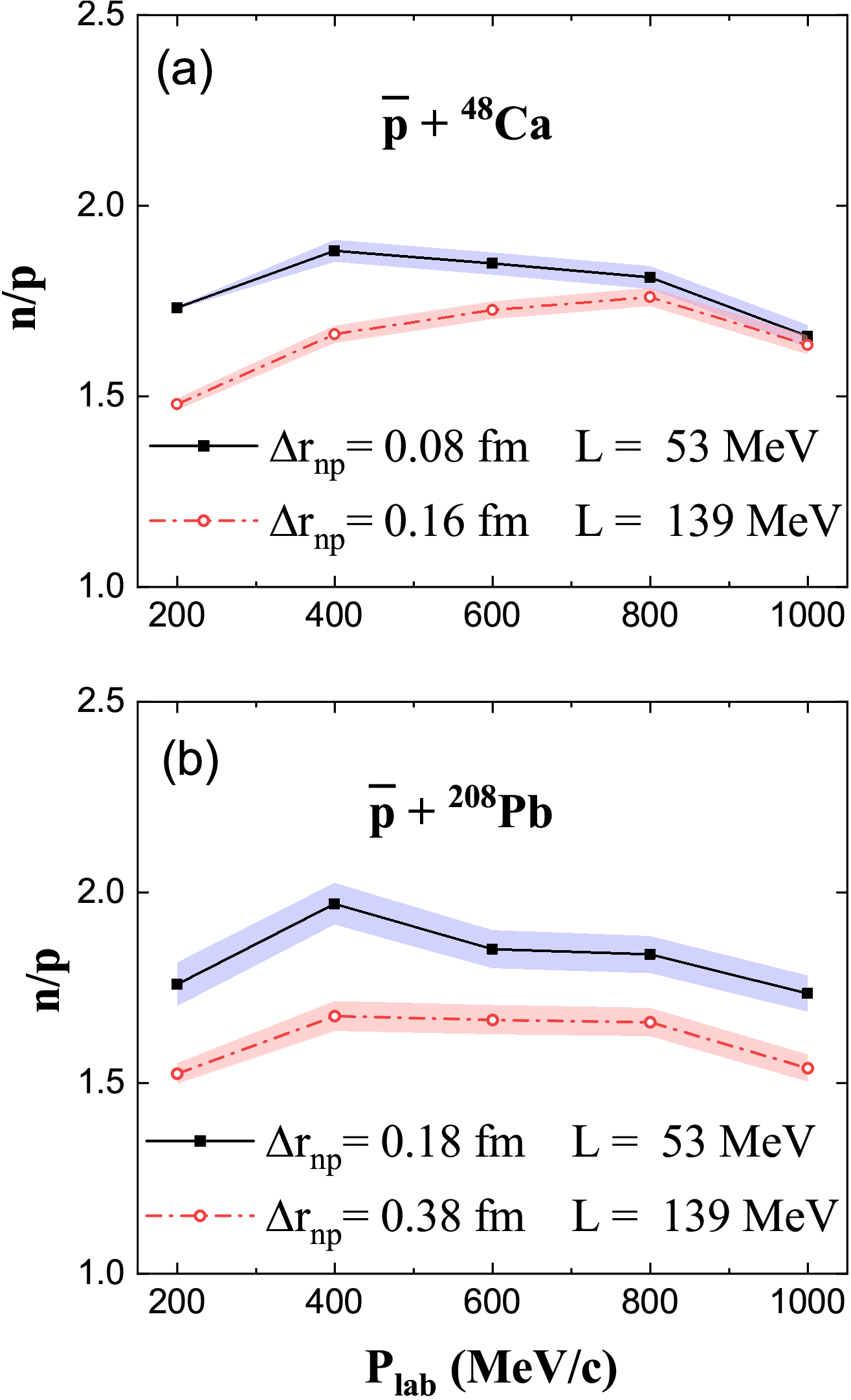}
\caption{\label{3} The beam momentum dependence of free neutron/proton ratio produced in collisions of (a) $\bar{\rm{p}} + ^{48}\rm{Ca}$ and (b) $\bar{\rm{p}} + ^{208}\rm{Pb}$ at the incident momentum of 200 MeV/c with the different neutron-skin thickness. }
\end{figure}

\begin{figure*}[ht]
\centering
\includegraphics[width=\linewidth]{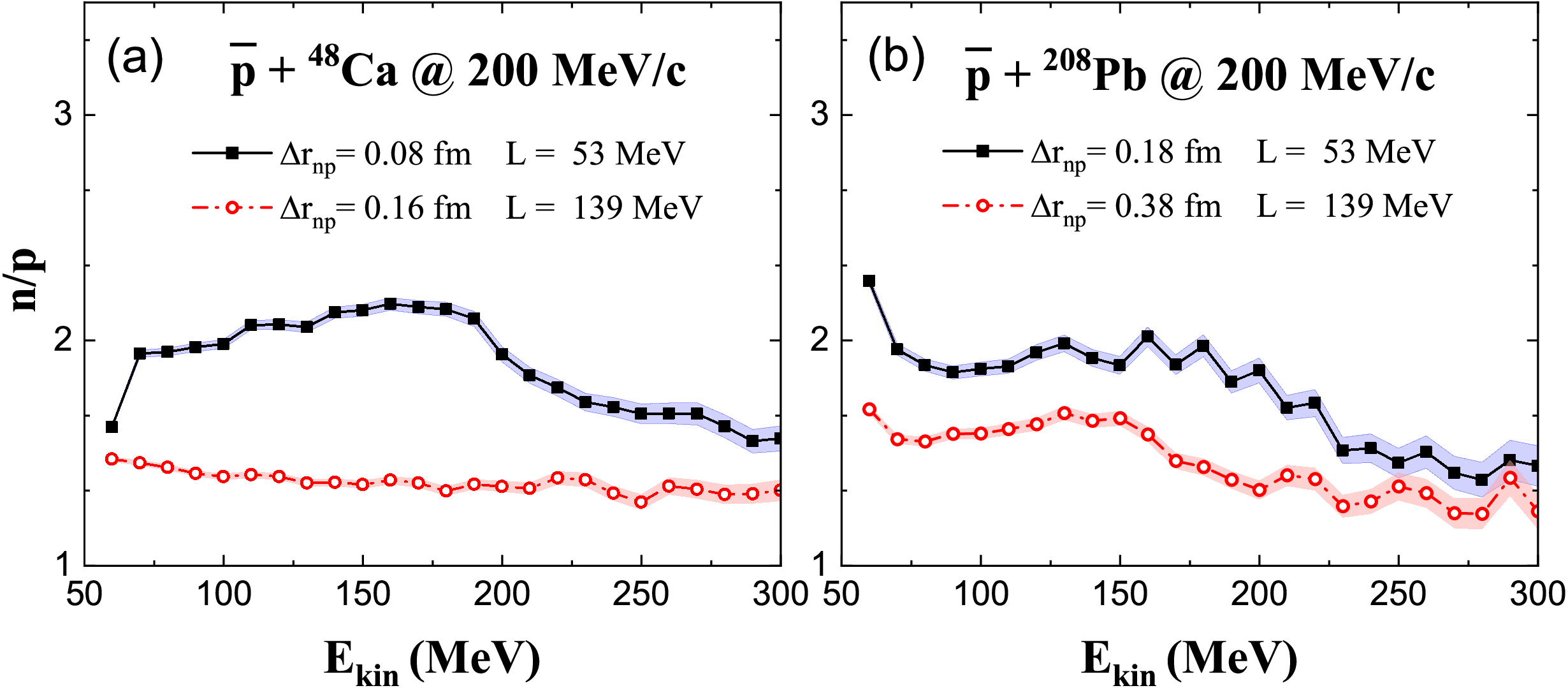}
\caption{\label{4}  The kinetic energy spectra of free neutron/proton ratio in the reactions of an antiproton on $^{48}\rm{Ca}$  (a) and $^{208}\rm{Pb}$ (b) nuclei at an incident momentum of 200 MeV.  }
\end{figure*}

In heavy-ion collisions, the spectra of the isospin ratios are related to the isospin-dependent in-medium cross sections, threshold energy correction, transportation dynamics etc. The emission of free nucleons, n/p ratio and pion spectra in the low-energy antiproton-induced reactions are associated with the antiproton-nucleon scatterings and low-density symmetry energy. The rescattering processes of pions and nucleons influence the n/p and $\pi^{-}/\pi^{+}$ spectra. Consequently, the n/p ratio with the kinetic energy above 50 MeV manifests the low-density symmetry energy information. The correlated investigation of the symmetry energy and the neutron-skin thickness is very necessary in the transport model. Shown in Fig. \ref{3} is the free n/p ratio as a function of the incident momentum of antiproton with different stiffness of symmetry energy. It is obvious that the symmetry energy effect is more pronounced at the lower incident momentum. The nucleons are the most copiously in the scattering of a pair of antiproton and nucleon. The symmetry potential enforces directly on the proton and neutron evolution. The low-energy antiproton prolongs the interaction of antiproton and nucleon, which enables the more pronounced isospin effect. In order to investigate the symmetry energy effects, we calculated the n/p ratios kinetic energy spectra for two neutron-rich nuclei, $^{48}\rm{Ca}$ and $^{208}\rm{Pb}$, at an incident momentum of 200 MeV/c for different neutron skin thicknesses. The findings are presented in Fig. \ref{4}. In two different reaction systems, it is observed that the n/p ratio is slightly enhanced with the softer symmetry energy compared to that with the stiffer ones. It is caused that the stiffer symmetry energy in the low-density region is characterized by a weaker repulsive potential for neutrons, whereas the softer symmetry energy features a stronger symmetry potential. Consequently, the more free neutrons are produced in the neutron-rich environment.

\begin{figure}[hp]
\centering
\includegraphics[width=0.85\linewidth]{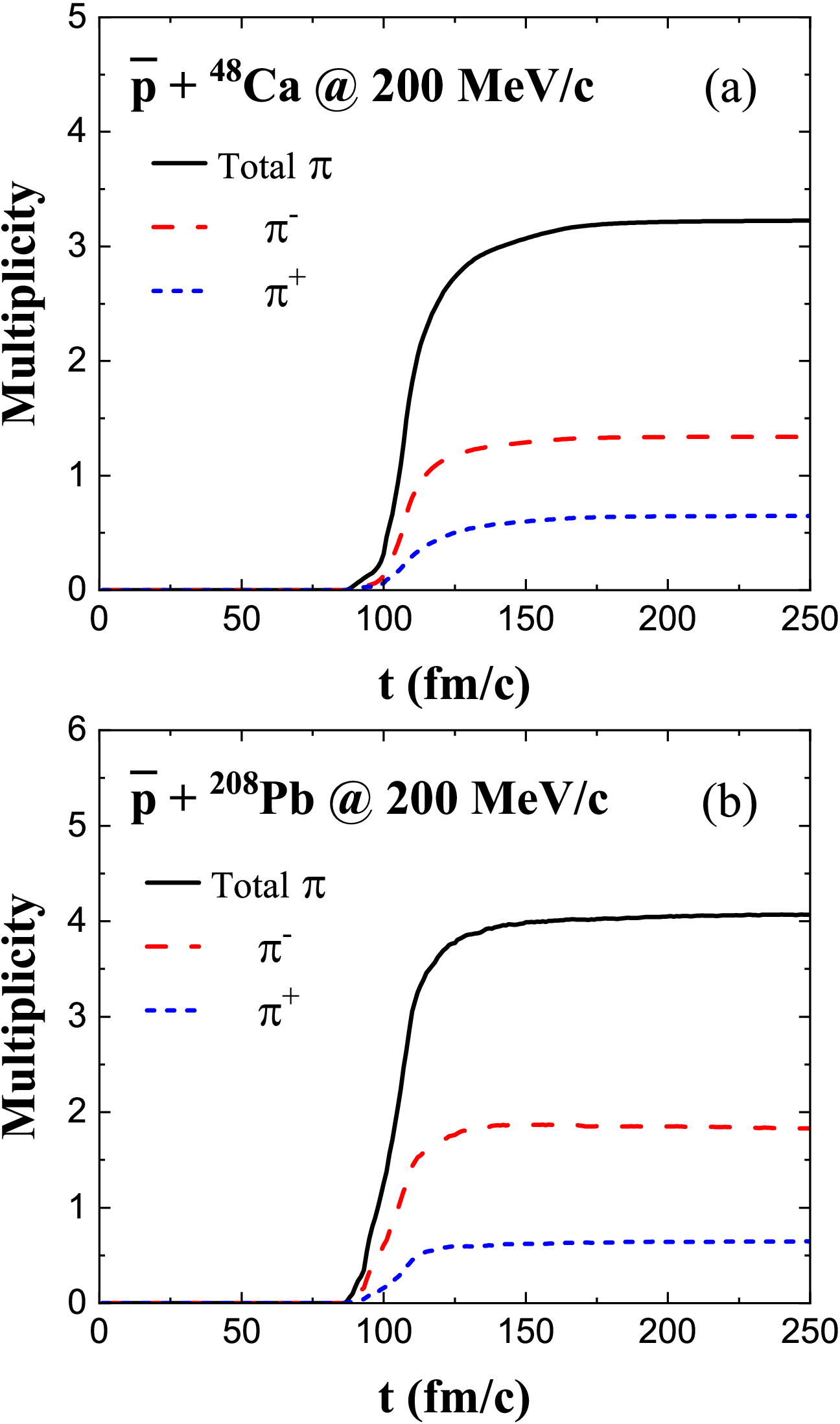}
\caption{\label{5} The temporal evolution of the pions produced in collision of $\bar{\rm{p}} + ^{48}\rm{Ca}$ (a) and $\bar{\rm{p}} + ^{208}\rm{Pb}$ (b) at 200 MeV/c. }
\end{figure}

\begin{figure}[h]
\centering
\includegraphics[width=\linewidth]{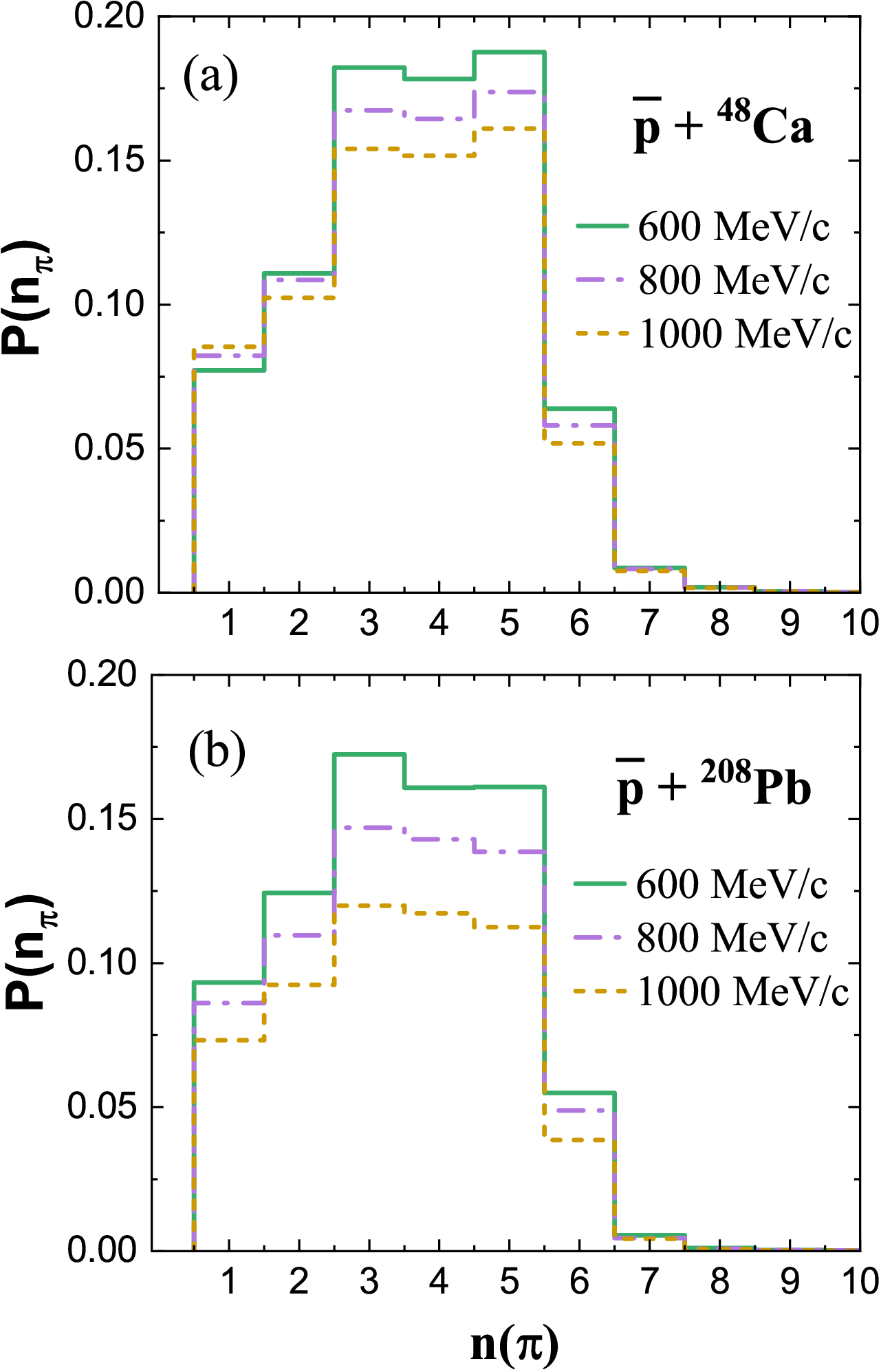}
\caption{\label{6} The probability distribution of the pion numbers from the annihilation of $\bar{\rm{p}} + ^{48}\rm{Ca}$ (a) and $\bar{\rm{p}} + ^{208}\rm{Pb}$ (b) at the incident momenta of 600, 800 and 1000 MeV/c, respectively. }
\end{figure}

\begin{figure}[h]
\centering
\includegraphics[width=\linewidth]{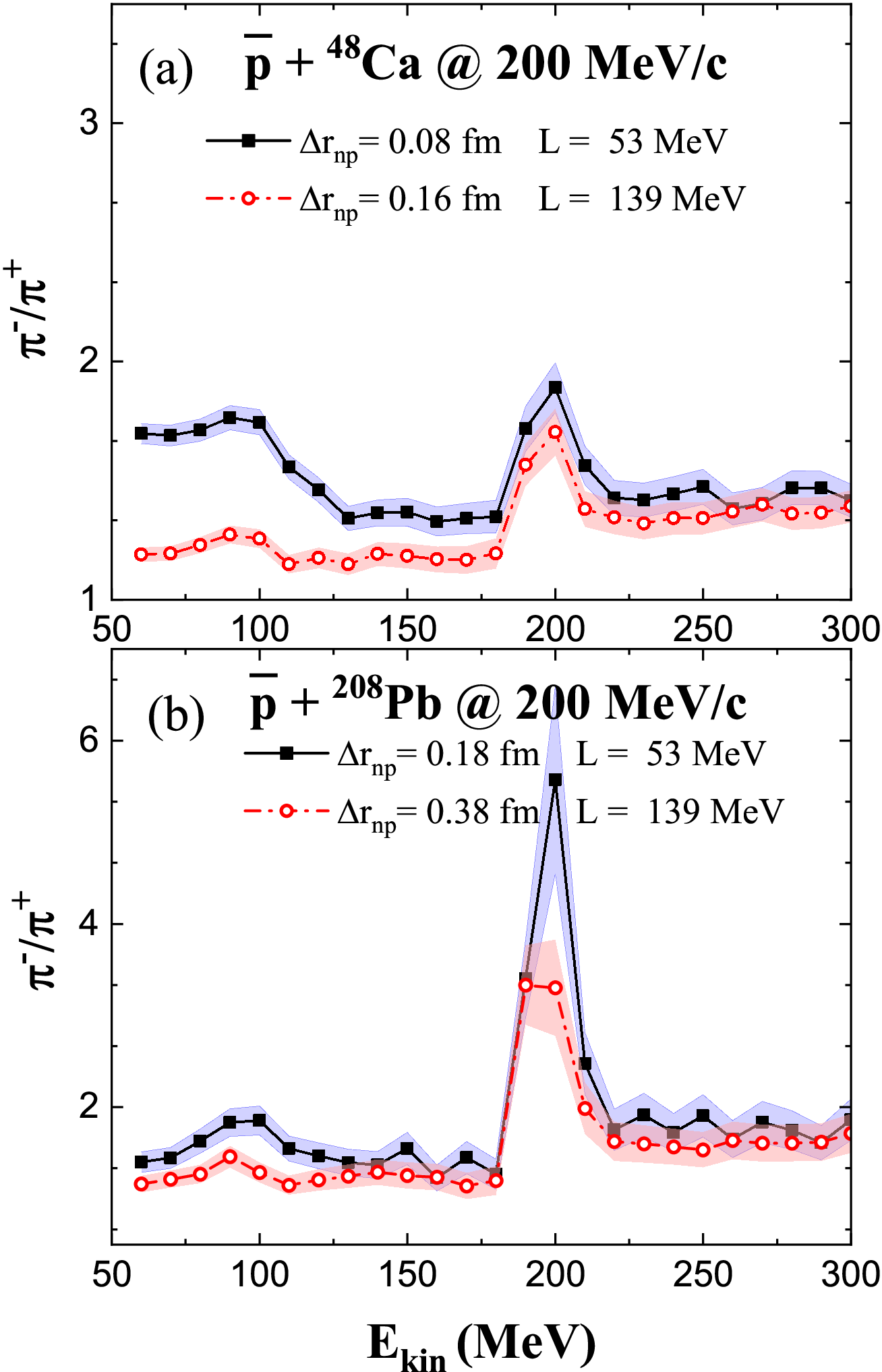}
\caption{\label{7} The kinetic energy spectra of $\pi^-$/$\pi^+$ ratio produced from the annihilation of an antiproton on $^{48}\rm{Ca}$  (a) and (b) $^{208}\rm{Pb}$ (b) nuclei at an incident momentum of 200 MeV/c. }
\end{figure}

The extraction of the high-density symmetry energy from the pion production in heavy-ion collisions has been extensively studied with the transport models, particularly the isospin-dependent Boltzmann-Uehling-Uhlenbeck (IBUU) and the relativistic Vlasov-Uehling-Uhlenbeck (RVUU) models \cite{Zhang17,Yong21,Fe06,Gao13}. The pion dynamics is complicated and associated with the pion-nucleon potential \cite{liu23}. We have investigated the production and temporal evolution of pions produced from the reactions of antiproton with $^{48}\rm{Ca}$ and $^{208}\rm{Pb}$ at incident momentum of 200 MeV/c. Fig. \ref{5} illustrates the temporal evolution of pion multiplicities for $\bar{\rm{p}} + ^{48}\rm{Ca}$  and $\bar{\rm{p}} + ^{208}\rm{Pb}$ collisions at 200 MeV/c. It can be observed that pion multiplicity in the antiproton reactions reaches the equilibrium at 150 fm/c. More pions are produced for the target $^{208}$Pb in comparison with $^{48}$Ca. The findings imply that the production of pions mainly originates from the annihilation process of an antiproton on the nuclear surface. Shown in Fig. \ref{6} is the multiplicity distribution of pions from the antiproton annihilation on $^{48}\rm{Ca}$ and $^{208}\rm{Pb}$ at incident momentum of 600 MeV/c, 800 MeV/c and 1 GeV/c, respectively. It can be seen that the pion production decreases with the incident momentum. The primordial pions are from the direct annihilation process, in which the multiplicities of 2-8 pions are produced \cite{Ho94}. The rescattering of pion and nucleon leads to the reduction of pion numbers and enables more complicated for the antiproton annihilation in nuclear medium. Studies from heavy-ion collisions have demonstrated that the $\pi^-/\pi^+$ ratio can be used as a sensitive probe for extracting symmetry energy information. In order to obtain more information about the symmetry energy from pion production, shown in Fig. \ref{7} is the kinetic energy spectrum of charged pions at an incident momentum of 200 MeV/c. It is interesting that the peak of $\pi^-/\pi^+$ corresponding the $\Delta(1232)$ resonance energy ($E_{\pi}$=0.19 GeV) is caused from the reabsorption of pion in nuclear medium. The soft symmetry energy leads to the larger $\pi^-/\pi^+$ ratio with the kinetic energy below 150 MeV, in particular for $^{48}$Ca. The reabsorption of pions by surrounding nucleons reduces the isospin effect for $^{208}$Pb.

\section{IV. Conclusions}

The neutron-skin thickness and its impact on the isospin observables in the low-energy antiproton induced reactions are thoroughly investigated within the LQMD transport model. The density dependence of symmetry energy is correlated with the neutron-skin thickness and
influences the isospin density transportation. The antiproton is mainly annihilated in the nucleus around the density of 0.6$\rho_{0}$ with the 3-5 pions per a pair of $\overline{p}p$ ($\overline{p}n$) annihilation. The soft symmetry energy with the slope parameter of L=53 MeV leads to the larger n/p ratio. The effect is obvious with decreasing the beam momentum in collisions of $\overline{p}+^{48}\rm{Ca}$ and $\overline{p}+^{208}\rm{Pb}$. The kinetic energy spectra of n/p ratios in the annihilation reactions are sensitive to the neutron-skin thickness above 50 MeV, in particular for the $^{208}\rm{Pb}$ target. The annihilation process proceeds the period of 50 fm/c for reaching the equilibrium of pion production, in which the symmetry potential enforces the nucleon evolution. The $\pi^-/\pi^+$ ratios are enhanced with the soft symmetry energy and a bump structure appears at the resonance energy ($E_{\pi}$=0.19 GeV). It is concluded that the symmetry energy effect is more pronounced in the antiproton collisions on $^{48}$Ca owing to the weak reabsorption reaction.

\textbf{Acknowledgements}
This work was supported by the National Natural Science Foundation of China (Projects No. 12175072 and No. 12311540139).

\end{document}